\documentclass[structabstract]{aa}

\usepackage{graphicx}
\usepackage{txfonts}
\usepackage{natbib}
\bibpunct{(}{)}{;}{a}{}{,} 

\begin{document}

 \title{A search for pulsations from the compact object of GRB 060218}

   \author{N. Mirabal\inst{1} \and E. V. Gotthelf\inst{2}
          }
\offprints{N. Mirabal, \email{mirabal@gae.ucm.es}}
\institute{Ram\'on y Cajal Fellow; Dpto. de F\'isica At\'omica, 
Molecular y Nuclear, Universidad Complutense de 
Madrid, Spain
\and Columbia Astrophysics Laboratory, Columbia University, New York,
NY 10027--6601, USA
           }

  \abstract
    {}
 {A fraction of massive stars are expected to collapse into 
compact objects (accreting black holes 
or rapidly rotating neutron stars) that successfully produce 
gamma-ray bursts (GRBs). 
We examine the possibility of directly observing these 
gamma-ray burst compact objects (GCOs)  using
post-explosion observations of past
 and future GRB sites.
}
{
We present a search for early 
pulsations from the nearby ($z$=0.0335) gamma-ray burst GRB 060218, which 
exhibited features possibly consistent with a rapidly spinning 
neutron star as its underlying GCO. We  also consider
alternative techniques that could potentially achieve a detection
of GCOs either in the Local Volume or near the plane of our
own Galaxy. 
}
 {
We report the non-detection of pulsations from the GCO of GRB 060218. 
In particular, fast fourier transform analysis applied to the 
light curve shows no significant power over the range of frequencies
$0.78 \ {\rm mHz} < f < 227 \ {\rm Hz}$ with an upper limit 
on the pulsed fraction of $\sim$ 2\%.
In addition, we present detection limits of current 
high-resolution archival X-ray images of galaxies 
within the Local Volume. The existing  
data could be harnessed to rule out the presence of any 
background contaminants at the GRB position of future nearby events.  
}
{The null detection of pulsations from the GCO of GRB 060218 
is most likely explained by the fact that the afterglow emission occurs
near the head of the jet and should be far removed
from the compact object. We also find that 
the comparison of pre- and post-explosion
explosion images of future GRBs within the Local Volume, as well as the
firm identification of a GCO within an ancient 
GRB remnant near the Galactic plane are extremely
challenging with current GeV/TeV capabilities. Finally, we conclude that
only under some very exceptional circumstances will it be possible
to directly detect the compact object responsible for gamma-ray bursts.
}

\keywords{Gamma rays: bursts -- Gamma rays: observations}

\headnote{}
\titlerunning{}
\authorrunning{Mirabal \& Gotthelf}

\maketitle

%

\section{Introduction}
Optical spectroscopy has verified the long-hypothesized origin of
gamma-ray bursts (GRBs) in the deaths of massive stars 
\cite[e.g.,][]{stanek1,hjorth}.
It is now accepted that ``long-duration'' ($>$ 2 s) GRBs accompany 
some core-collapse supernovae of Type Ic, in which a 
compact object produced during the collapse of a Wolf-Rayet progenitor
powers a focused jet that bores its
way through the stellar envelope \citep{woosley2}. Unfortunately,
the exact nature and eventual fate of the gamma-ray burst compact object (GCO) 
responsible for the the launch of
the GRB jet remains unknown. The most promising alternatives include 
highly magnetized neutron stars \citep{wheeler,thompson} and 
accreting black holes characterized as ``collapsars'' \citep{woosley,macf}.
The detection of such GCOs would not only open a new frontier in
our understanding of compact object formation, 
but it would help us address the circumstances that give rise 
to a variety of stellar explosions. 

Much of the effort aimed at revealing the nature of 
GCOs has centered on the possible interaction of the
GRB with its surroundings. \citet{brown}
suggested that GRB explosions could
lead to the formation of X-ray binaries with a black hole primary.
The identification of
GRB remnants in our Galaxy or in nearby galaxies has also been proposed
as a possible way to  pinpoint the
GCO responsible for the explosion \citep{perna, ayal,ramirez}. 
Looking for the compact object produced
at the site of historical core-collapse supernovae has also been addressed
by \citet{perna1}.
Using data from the Burst and Transient Experiment (BATSE) onboard the
Compton Gamma Ray Observatory (CGRO),
\citet{macbreen} inspected cumulative 
light curves of GRBs for possible signatures of a black hole being
either spun up or down during the accretion process. Other surveys have
focused on the search for gravitational waves associated with GRBs 
\citep{abbott}.
Most recently, \citet{mazzali} have used spectral analysis of
the broad-lined supernova SN 2006aj to 
infer the formation of a magnetar associated with
GRB 060218. While these studies provide important insight into 
the origins of the explosion, 
further observational work is required to detect and distinguish 
GCOs directly.

Theoretical studies of GRBs suggest that the 
combination of rapid rotation and large magnetic fields in
neutron stars ({\it magnetars}) 
could generate an energy of $10^{51}$--$10^{52}$ erg on a 
10--100~s timescale required to power a long-duration GRB 
\citep{thompson}. 
A direct detection of pulsations at 
the site of a GRB would provide strong support for such hypothesized 
neutron star at the center of the explosion.  
So far, searches for pulsation signatures 
have found no significant evidence of periodicity in the prompt 
gamma-ray (25--320 keV energy range) light curves of GRBs \citep{deng}. 
However, very recent observations
have produced tantalizing but controversial evidence of quasi-periodic
variations in the gamma-ray light curve of GRB 090709A \citep{mark,
gole,gotz,mirabal3}. 

Owing to its close proximity ($z$=0.0335) and the 
early detection of an associated supernova with the {\it Swift} satellite,
the observations of GRB 060218/SN 2006aj \citep{campana,mirabal2,
modjaz,pian,soderberg} constitute an ideal dataset to
search for pulsations following the GRB onset. 
Additional motivation for a thorough pulsation search is provided by 
spectral modeling of its associated supernova SN 2006aj suggesting a 
nascent neutron star as the likely GCO  for GRB 060218 \citep{mazzali}.

In this paper, we present an early search for pulsations from the
GCO of nearby GRB 060218, and an analysis of alternative
observations that may produce a direct view of the GCO in the future. 
The paper is organized as follow. In Sect. 2, we describe a search for
coherent pulsations from the GRB 060218. 
Sect. 3 discusses our results and examines potential 
techniques that could reveal the nature of GCOS. Finally,
conclusions are presented in Sect. 4.
Throughout we adopt a 
concordance cosmology model 
($H_{\rm 0} = 71$ km s$^{-1}$ Mpc$^{-1}$, $\Omega_{\rm{m}}=0.27$
and $\Omega_\Lambda = 0.73$).

\section{Observations and Data Analysis}

Timing observations of GRB 060218 were obtained with the X-ray 
telescope (XRT; Gehrels et al. 2004) on-board the {\it Swift}
satellite. The XRT focal plane sensor is a $600 \times 602$ pixel CCD
sensitive to photons in the 0.2--10 keV energy range. Data were
acquired in Windowed Timing (WT) mode starting 159~s after the burst.
This mode provides 1.77 ms time resolution by rapidly clocking out the
central 200 columns in the column direction with the resulting loss of
imaging in one spatial dimension. We analyzed standard pipeline
products created using processing version v3.9.10 made available in
the {\it Swift} archive \footnote{See
http://heasarc.gsfc.nasa.gov/docs/swift/archive/}.

\begin{figure}
\centering\resizebox{7.0cm}{!}{\includegraphics{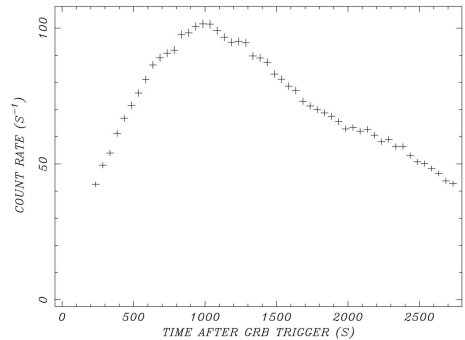}}
  \caption{
Initial {\it Swift} XRT light curve of GRB 060218 acquired 159~s after 
the GRB trigger, in the 
0.2--10 keV energy band.}
  \label{light}
\end{figure}

Data collected in the first orbit spanned an initial 2540~s with a
mean count rate of 72 counts s$^{-1}$ in a $1'$~wide extraction
aperture centered on the source. The light curve rises to a peak 100
counts s$^{-1}$ at the 800~s mark before slowly falling back to the
initial detected rate of 30 counts s$^{-1}$. In the next interval, following 
a 3160 s spacecraft orbit gap due to Earth block, 
the source is barely detectable (mean rate of 0.7 counts s$^{-1}$ over
120~s), then nearly undetectable during 22 $\times$ 40s
orbit-to-orbit monitoring intervals over 35~hrs, for a total of 808~s
of exposure, resulting in 32 source counts and 8 background
counts. Apart from the highly variable source flux rate, the
background remained constant during all observation intervals.
A light curve of the {\it Swift} events for the first interval following
detection is shown is Fig. \ref{light};
note that count rates for the later observations are essential zero on this
scale.

\begin{figure}
\centering\resizebox{7.0cm}{!}{\includegraphics{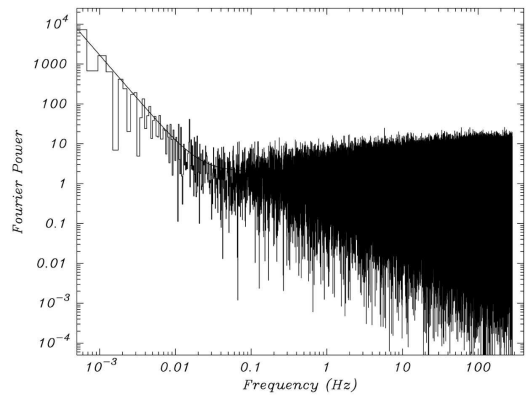}}
  \caption{
Power spectrum of GRB 060218. The spectrum shows
no significant coherent signal and it is dominated by red noise
at low frequencies $\lesssim 0.03$~Hz.}
  \label{fft}
\end{figure}

We searched for a coherent signal in the initial {\it Swift} XRT data
of GRB 060218 using a fast Fourier transform (FTT). The
barycenter-corrected data taken during the first orbit was re-binned
in 1.77~ms steps and transformed using a $N=2^{21}$ element FFT.  No
significant power is found over the range of search frequencies $0.78 \ {\rm
mHz} < f < 227 \ {\rm Hz}$ searched. The resulting power spectrum for
GRB 060218 is displayed in Fig. \ref{fft} and shows characteristics of
both white (power independent in frequency, mean of 2) and red noise
(power modeled by a power law with best fit index $\Gamma = -2.2$),
the latter dominating at frequencies above 30~mHz.  In the white noise
regime, the maximum FFT power is $S = 25.67$ at 223~Hz
($P=4.48$~ms). This power is consistent with random fluctuations,
corresponds to a false detection probability of $\wp =
N_{trials}\exp^{-S/2} = 2^{21} \times 2.6 \times 10^{-6} > 1$ for a
coherence sinusoidal.  Formally, the corresponding pulsed fraction
limit is $f = 1.8\%$ ($ =
\sqrt{2S/N}$, $N = 162$~k detected photons), with 50\% probability.
For periods in the magnetar range ($0.01 \lesssim f \lesssim 1$~Hz),
which partially overlap the red noise regime, the peak FFT power is $S
= 37.51$ after allowing for the modeled red noise contribution,
corresponding to a blind search false positive detection probability
of $\wp = 0.015$ ($\approx 2.5 \sigma$) and pulsed fraction limit of
$f = 2.1\%$. Similar results are obtained using the Rayleigh, or
$S=Z^2_1$ test, where the power is distributed like $\chi^2$ with
two degrees of freedom, as for the FFT power.

To look for possible non-stationary oscillations, a similar timing
analysis was done on 100-s intervals using the burst and no
significant signal is found to a pulsed fraction limit of ~10\%,
(maximum FFT power~$S \approx 10, N \approx 4000$ counts), with 50\%
probability.

\section{Discussion}

In a different astrophysical context, the absence of coherent pulsations 
from a newly formed compact object would be intriguing. 
In the case of GRBs; however, the absence of coherent pulsation
is not totally unexpected. The reason is that the 
observed afterglow/breakout emission most likely 
occurs near the jet head far removed
from the central engine \citep{zhang1}. As a result, 
temporal and spectral signatures (if any) 
associated with the GCO are probably swamped by the 
afterglow emission \citep{mirabal1,sako, butler}. 

Even in the absence of a bright afterglow, as is the case
for GRB 060218, it is not entirely clear that the temporal signatures will
be retained during the expansion of the jet. 
Consequently, we argue that early searches for pulsations in GRBs 
will not always provide clues about the nature of GCOs.
We note, however, that our results do not necessarily rule out neutron star
models for GCOs. In fact, the range of 
frequencies considered here may only rule
out millisecond pulsars with relative slow spin periods $f < 283$ Hz. 
As argued by \citet{thompson}, 
an energy release $E_{\gamma} \sim 10^{50}$ erg implies that
the spin period of the neutron star at birth might be as low 
as 0.1 ms. 

Interestingly, our failed search forces us to consider 
alternative approaches for observing GCOs directly 
such as direct comparison of pre- and post-explosion
X-ray images of future nearby GRB localizations 
\citep{pian2,kouveliotou} or  the firm identification 
of a GCO within the interior of a GRB remnant in our own Galaxy \citep{ioka}. 
In the case of pre- and post-explosion comparisons, the appearance of 
an X-ray source at the GRB localization,
once the early GRB afterglow and underlying 
supernova have significantly subsided \citep{kouveliotou}, would provide
direct access to the GCO itself. 

If GRBs originate in binary systems
and the binary remains bound after the explosion \citep{davies}, it is possible
that a soft X-ray transient will be formed after the explosion 
via black hole accretion \citep{brown}; or through accretion onto 
a neutron star from the stellar wind of
an O or B companion \citep{lewin}. Since GRBs 
are a subset of supernovae that in
some cases may give birth to rapidly rotating
magnetars, it makes sense to look for X-rays
from an isolated rotation powered pulsar, not
just an accreting binary. \citet{perna} has computed the expected X-ray 
luminosity as a function of time for neutron stars born with
different magnetic fields and spin periods. 
While the detection of a steady X-ray source at the site of the explosion
would not by itself distinguish between a black hole or neutron star
compact object, it would provide direct access to the compact 
object underlying the explosion. 

\begin{figure}
  \resizebox{\hsize}{!}{\includegraphics{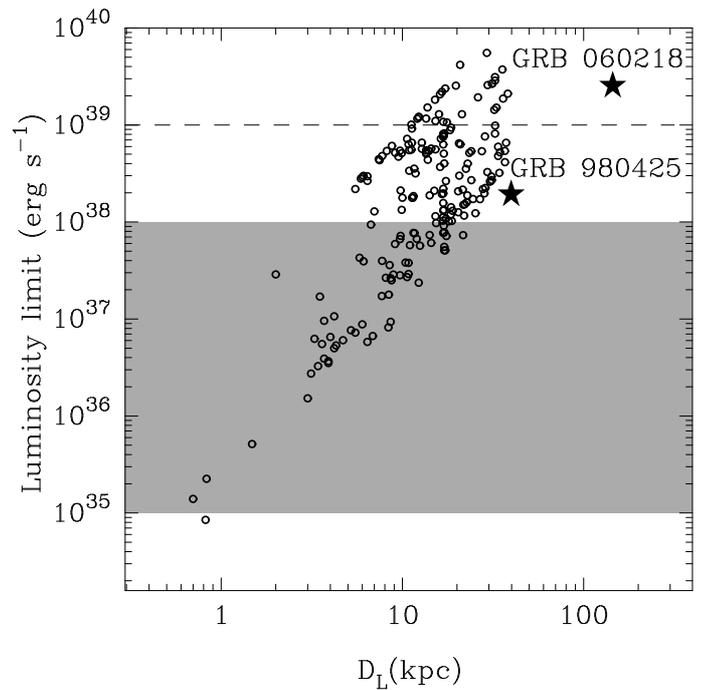}}
  \caption{
X-ray luminosity limits in the 0.3--10 keV band for the subsample of galaxies
tabulated in \citep{liu} plotted as a function of distance $D_{\rm L}$ 
(open circles).
For comparison we also show the respective limits for
GRB 980425 \citep{galama} and GRB 060218 (stars) assuming
an X-ray flux limit of $f_{X}=10^{-15}$ ergs~cm$^{-2}$ ~s$^{-1}$.
The shaded region shows the expected range in X-ray luminosities for
X-ray binaries \citep{grimm}. The dashed line near the top
denotes the Eddington limit for a stellar mass black hole.}
  \label{lum}
\end{figure}

Apart from modeling the SN and afterglow evolution, the main
challenge with pre- and post-explosion comparisons is 
ruling out unrelated background sources 
such as AGN and pre-existing X-ray binaries sources at the GRB site. 
Fortunately, there already exists an extensive high-resolution 
X-ray image archive of nearby 
galaxies acquired during the past decade \citep{liu}.  
The recent compilation by \citet{liu}
includes 383 galaxies within 40 Mpc. After excluding early-type galaxies,
this leaves a subset of 200 late-type, irregular, and peculiar 
galaxies within 40 Mpc. 
Potentially, this extensive X-ray sample can be used to 
to set restrictive upper limits on the X-ray emission at the GRB site 
prior to the explosion. Fig. \ref{lum} shows the X-ray luminosity limits L$_{X}$    
for 200 galaxies listed in \citet{liu}  as a function of distance. 
Overplotted in Fig. \ref{lum}
are the luminosity limits reachable for nearby GRB 980425 \citep{galama} 
and GRB 060218 with current X-ray telescopes assuming
an X-ray flux limit 
$f_{X} \sim 10^{-15}$ ergs cm$^{-2}$ s$^{-1}$. In the case of GRB 060218, 
a detection of a newly formed X-ray binary would have to exceed the levels of 
Eddington luminosity for a
stellar-mass black hole. The situation improves for less distant
events. 

Barring the highly unlikely explosion of a GRB in our Galactic neighborhood
\citep{stanek2}, a direct detection of a GCO embedded in a 
GRB remnant near the Galactic plane
will have to rely heavily on studies based on 
morphology alone \citep{perna,ioka}. 
Numerical simulations suggest that the evolution of the GRB explosion would 
create non-spherical topologies that could stand out against the
remnants of other stellar explosions
\citep{ayal,ramirez2}. However,
the stability of such non-spherical morphology has been called into
question given the sideways evolution of the relativistic 
jet \citep{zhang2}. Indeed,
GRB remnants may lurk within the 
numerous shells identified near the Galactic plane,
 but distinguishing 
them from other types of stellar remnants will be a very tall task
\citep{daigle}. 
In practical terms,  the actual probability of GRB remnant 
detection is determined simply by the available volume that can be sampled and
accurately classified. Currently, the likelihood of detecting such a GRB 
remnant with the available sample of high-energy sources  
appears limited \citep{aha1,holder,rico,abdo}.  Fortunately, future
dedicated GeV/TeV surveys could bring such detection within reach
\citep{buckley}.

\section{Conclusions}

In this paper we have conducted a search for early pulsations from
the compact object of the nearby GRB 060218. No
significant evidence of pulsations is observed. 
This null detection
suggests that the afterglow continuum most likely swamps any 
pulsation/modulation signature or that the viewing angle into the
progetitor is rather unfavorable. By itself, this result
is not sufficient to rule out neutron star models. 
It should be noted that high redshifts and bright
relativistic jets will challenge a sensitive search for pulsations in
other GRBs detected by {\it Swift}.

Based on our analysis, 
opportunities of observing GCOs directly either through their interaction with
a bound stellar binary companion or via the detection of a compact object 
embedded within
a GRB remnant near the Galactic plane offer modest hope for 
moving beyond afterglow science in the near future. 
The situation could
radically change with a reasonably placed GRB in the Local Volume 
(\emph{e.g.}, SN 1987A) or 
through substantial progress in our GeV/TeV capabilities. 
Thus, we conclude that observational progress on the nature of GCOs 
will once again have to rely heavily 
on the serendipity that is central to the origin and history
of the GRB field.

\begin{acknowledgements}
We thank Jules Halpern for stimulating conversations. N.M. acknowledges 
support from the Spanish Ministry of Science
and Technology through a Ram\'on y Cajal fellowship.

\end{acknowledgements}

{}

\bibliographystyle{aa} 

\end{document}